\documentclass[twocolumn,english,pra,showpacs,floatfix]{revtex4}
\usepackage[T1]{fontenc}
\usepackage[latin1]{inputenc}
\usepackage{graphicx}
\usepackage{amssymb}

\makeatletter

\usepackage{graphicx}

\usepackage{babel}
\makeatother

\newcommand \bea{\begin{eqnarray}}
\newcommand \eea{\end{eqnarray}}

\newcommand{\av}[1]{\langle{#1}\rangle}

\begin{document}
\title{Itinerant Ferromagnetism in ultracold Fermi gases}
\author{H. Heiselberg}
\affiliation{Applied Research, DALO, Lautrupbjerg 1-5, DK-2750 Ballerup, Denmark}

\begin{abstract}
Itinerant ferromagnetism in cold Fermi gases with repulsive interactions is studied 
applying the Jastrow-Slater approximation generalized to finite polarization
and temperature. For two components at zero temperature a second order transition is found
at $ak_F\simeq0.90$ compatible with QMC.
Thermodynamic functions and observables such as the compressibility and spin susceptibility 
and the resulting fluctuations in number and spin are calculated.
For trapped gases the resulting cloud radii and kinetic energies are calculated
and compared to recent experiments. Spin polarized systems are recommended for effective
separation of large ferromagnetic domains. Collective modes are predicted and
tri-critical points are calculated for multi-component systems.
\end{abstract}
\maketitle

\section{Introduction}

Ultracold Fermi systems with strong attraction between atoms has
led to important discoveries as universal physics and the BCS-BEC
crossover.  Recently strongly repulsive interactions has been studied
and a transition to a ferromagnetic phase was observed in the
experiments of Jo et al. \cite{Jo}. Earlier Bourdel et al. \cite{Bourdel} and
Gupta et al. \cite{Gupta} also observed a transition when the
interactions became strongly repulsive near Feshbach resonances.  A
phase transition from a paramagnetic (PM) to ferromagnetic (FM) phase
was predicted long ago by Stoner \cite{Stoner} based on the
Hartree-Foch mean field energy and has recently been confirmed by more
elaborate calculations including fluctuations \cite{Duine,Conduit} and by
QMC \cite{Pilati,Chang}.  The calculated transition points and
order of the transition differ also
from experiment \cite{Jo}.  The FM transition is disputed by Zhai
\cite{Zhai} who claims that the experimental data is compatible
with strongly correlated repulsive Fermi systems which would explain
the inability to observe FM domains in Ref. \cite{Jo}.

It the purpose of this work to clarify the phase diagram of strongly
repulsive Fermi atomic systems as well as to calculate thermodynamic
functions and measurable observables in atomic traps that clearly can
distinguish the FM and PM phases and determine the order of the
transition and the universal functions.  By extending the
Jastrow-Slater model \cite{Vijay,long,Cowell} to finite
polarization and temperature, we calculate the free energy and find a
second order FM transition in a repulsive Fermi gas. A number of
thermodynamic functions as the spin susceptibility, compressibility,
and observables as radii and kinetic energies can be compared to
experiments, and others as fluctuations, collective oscillations and
phase separation can be predicted.

As a start the dilute limit model of Stoner is extended to finite
temperature and the polarization and order of the transition is determined
and compared to second order calculations. Subsequently, 
the Jastrow-Slater approximation is described for the correlated manybody
wave-function in the strongly interacting limit and extended to
finite polarization and temperature. Detailed calculations of the free energy and 
a number of thermodynamic functions are given. In particular the
spin-susceptibility and compressibility are used for calculating fluctuations
in spin and total particle number in section III. In section IV finite traps
are considered and the cloud radii and kinetic energies are calculated and
compared to recent
experiments \cite{Jo}. Collective modes are discussed in section V. 
Multi-components systems are discussed in section VI and a new string of critical
points is found and plotted in a multi-component phase diagram. 
Finally, a summary and outlook is given.

\section{Ferromagnetic transition}

The models for repulsive ultracold Fermi gases in Refs.
\cite{Stoner,Duine,Conduit,Pilati,Chang} all predict a phase transition somewhere
near the unitarity limit $ak_F\sim 1$ but the phase diagrams disagree
quantitatively as well as qualitatively concerning the order and critical points.

For a reference model we start with a simple finite temperature
extension of the Hartree-Fock approximation originally studied by
Stoner \cite{Stoner}, which is a dilute
limit expansion to first order in the scattering length. 
Subsequently, we calculate the phase diagram in the JS
approximation and compare to those in the dilute limit to first
\cite{Stoner} and second \cite{Duine,Conduit} order as well as QMC
\cite{Pilati,Chang}.

\subsection{Dilute approximations}

\begin{figure}
\includegraphics[scale=0.6,angle=0]{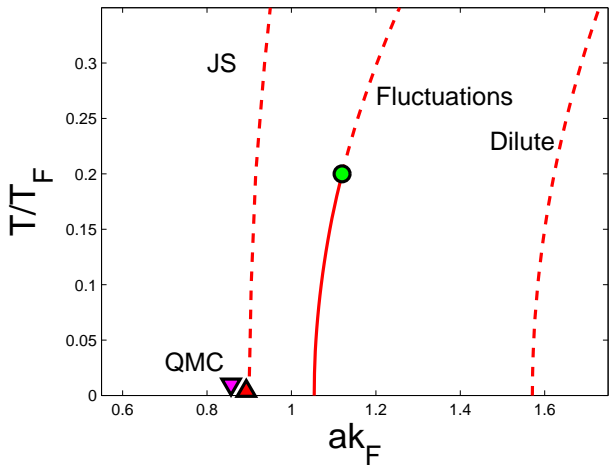}
\vspace{-0.5cm}
\caption{Phase diagram for a two-component Fermi gas with repulsive interactions.
Full (dashed) curves indicate first (second) order PM to FM transitions within JS and 
dilute approximations with \cite{Conduit} and without 
\cite{Stoner} fluctuations. The circle indicates a tri-critical point. Triangles show the QMC transition points at zero temperature of Refs. \cite{Pilati,Chang}.}
\end{figure}

A dilute ($k_Fa\ll1$) degenerate Fermi gas with atoms in spin states $\sigma=1,..,\nu$ with densities $\rho_\sigma=k_{F,\sigma}^3/6\pi^2$ and Fermi energy 
$E_{F,\sigma}=k_BT_{F,\sigma}=\hbar^2 k_{F,\sigma}^2/2m$ has the free energy
\bea \label{dilute}
 f = \frac{3}{5}\sum_{\sigma} E_{F,\sigma} \rho_\sigma 
     \, +    \sum_{\sigma<\sigma'}\frac{4\pi a}{m}\rho_\sigma \rho_{\sigma'}  \, + f_T \, .
\eea
It consists of the kinetic energy, the interaction energy to lowest order in the scattering length $a$
as in the Stoner model \cite{Stoner},
and the thermal energy $f_T=-(\pi^2/4)(m^*/m)\sum_\sigma \rho_\sigma T^2/E_{F,\sigma}$
at low temperatures $T\ll E_{F,\sigma}$.
In the dilute limit 
the effective mass $m^*/m=1+[8(7\ln 2-1)/15\pi^2]a^2k_F^2$ in two-component symmetric systems
only deviates from the bare mass to second order in the interaction parameter $ak_F$.

The density of the components 
are equal only in the PM phase and when the components are balanced initially.
In the following we define an average Fermi wave number $k_F$ from the total density
$\rho=\nu k_{F}^3/6\pi^2$.

We postpone multicomponent systems to sec. VI and concentrate first on
two spin states, e.g. $\sigma=\uparrow,\downarrow$ with total density
$\rho=\rho_\downarrow+\rho_\uparrow$. The population of spin states are allowed to change (polarize) in order to observe phase transitions to itinerant ferromagnetism.
The polarization (or magnetization) $P=(\rho_\downarrow-\rho_\uparrow)/\rho$ of the ground state phase is found by minimizing the free energy at zero magnetic field. 
The free energy of a low temperature ideal gas is $f_0=E_F\rho(3/5-\pi^2T^2/4E_F^2)$. 
Expanding Eq. (\ref{dilute})
for small polarization leads to a Ginzburg-Landau type equation for the free energy
\bea \label{GL}
 \frac{f}{f_0} &=&1+\frac{10}{9\pi}ak_F+\frac{5}{9} \left[\frac{\chi_0}{\chi_T} P^2 
    +\frac{1}{27}P^4\right] \, ,
\eea
to leading orders in interaction, polarization and temperature.
Here $\chi_T=(\partial^2 f/\partial P^2)^{-1}$ is the isothermal spin-susceptibility
given by
\bea \label{chiStoner}
  \frac{\chi_0}{\chi_T} = 1-\frac{2}{\pi}ak_F+\frac{\pi^2}{12}\frac{T^2}{T_F^2} \, ,
\eea
where $\chi_0=3\rho/2E_F$ is the spin-susceptibility for an ideal gas at zero temperature.
$\chi_T$ becomes singular when 
\bea \label{adilute}
   ak_F=\frac{\pi}{2}\left(1+\frac{\pi^2}{12}\frac{T^2}{T_F^2}\right) \, ,
\eea
where the free energy of Eq. (\ref{GL}) predicts a second order phase transition 
from a PM to a FM (see Fig. 1) in accordance with the zero temperature result of Stoner \cite{Stoner}. The polarization is
$P=\pm\sqrt{27(ak_F/\pi-1/2)}$ at zero temperature but
quickly leads to a locally fully polarized system $P=\pm1$ due to the small fourth order
coefficient in Eq. (\ref{GL}).

However, the predicted transition occurs close to the unitarity
limit where the dilute equation of state Eq. (\ref{dilute}) is not valid. Higher orders are important
as exemplified by including fluctuations, i.e. the next order $a^2$ correction. 
As found in Refs. \cite{Duine,Conduit} fluctuations
change the transition from second to first order at low
temperatures up to a tri-critical point at temperature $\simeq 0.2T_F$, where the
transition becomes second order again (see Fig. 1). 
However, the 2nd order expansion is not valid either in the unitarity limit.

\subsection{Jastrow-Slater approximation}

The JS approximation applies to both strongly attractive and repulsive crossovers where it already has
proven to be quite accurate for predicting universal functions and parameters. 
The JS approximation is the lowest order in a constrained variational (LOCV)
approach to calculate the ground state energies of strongly correlated systems.
It was developed for strongly interacting and correlated
Bose and Fermi fluids respectively such
as $^4$He, $^3$He and nuclear matter \cite{Vijay}.
JS was among the earliest models applied to the unitarity limit and
crossover of ultracold Fermi \cite{long} and Bose \cite{Cowell} atomic gases.
As explained in \cite{Vijay,long,Cowell} the JS wave function 
\begin{eqnarray}
\Psi_{JS}({\bf r}_1,...,{\bf r}_N)= 
\Phi_S\prod_{i,j'}\phi({\bf r}_i-{\bf r}_{j'}) \,,
\end{eqnarray} 
incorporates essential two-body correlations in the Jastrow function $\phi(r)$.
The antisymmetric Slater wave function $\Phi_S$ for free fermions
$\Phi_S$ insures that same spins are spatially anti-symmetric. 
The Jastrow wave function only applies to particles with different spins
(indicated by the primes). The pair
correlation function $\phi(r)$ can be determined variationally by
minimizing the expectation value of the energy, $E/N = \langle \Psi
|H| \Psi \rangle \ /\ \langle \Psi | \Psi \rangle$, which may be
calculated by Monte Carlo methods \cite{Casulleras,Chang}.
At distances shorter than the interparticle spacing two-body clusters dominate and
the Jastrow wave function $\phi(r)$ obeys the
Schr\"odinger equation for a pair of particles of different spins
interacting through a potential $U(r)$ 
\bea \label{Schrodinger}
  \left[-\frac{\hbar^2}{m}\frac{d^2}{dr^2} +U(r)\right] r\phi(r) 
 =2\lambda\, r\phi(r)
   \, ,
\eea
where the eigenvalue is the interaction energy of one atom $\lambda=E_{int}/N$.
Most importantly, the boundary condition at short distances ($r=0$)  is
given by the scattering length 
\bea \label{a}
\frac{(r\phi)'}{r\phi} =-\frac{1}{a} \, .
\eea
Many-body effects become important
when $r$ is comparable to the interparticle distance $\sim k_F^{-1}$, but are found to be small
\cite{Vijay,long,Cowell}. Here the boundary conditions
that $\phi(r>d_\sigma)$ is constant and $\phi^{\prime}(r=d_\sigma)=0$ 
are imposed at the healing distance $d_\sigma$, 
which is determined self consistently from number conservation
\bea \label{numberspin}
  (\rho-\rho_\sigma)\int_0^{d_\sigma} \frac{\phi^2(r)}{\phi^2(d_\sigma)} 4\pi r^2dr = 1 .
\eea
The prefactor $\rho-\rho_\sigma=\rho_{\sigma'}$ takes into account 
that a given spin $\sigma$ only interacts
and correlates with unlike spins $\sigma'\ne\sigma$.
In the dilute limit $\phi(r)\simeq 1$ and so $d_\sigma=(9\pi/2)^{1/3}k_{F,\sigma'}^{-1}$.
In the unitary limit $a\to \pm\infty$ 
the healing length approaches
$d_\sigma=(3\pi)^{1/3}k_{F,\sigma'}^{-1}$ in stead.
Generally the healing length is of order the Fermi wavelength of the other component,
$d_\sigma k_{F,\sigma'}\sim 1$.

For a positive scattering length the interaction energy $\lambda$ is positive and 
the solution to Eq. (\ref{Schrodinger}) is 
$r\phi(r)\propto\sin[k(r-b)]$ with $\lambda_\sigma=\hbar^2k^2/2m$. Defining $\kappa_\sigma=kd_\sigma$
the boundary conditions and number conservation requires \cite{Cowell}
\bea \label{pos}
 \frac{a}{d_\sigma} = \frac{\kappa_\sigma^{-1}\tan\kappa_\sigma-1}{1+\kappa_\sigma\tan\kappa_\sigma} \,.
\eea
The resulting interaction energy reproduces the correct dilute limit result of Eq. (\ref{dilute}).
In the unitarity limit $a\to +\infty$, the positive energy
solution reduces to $\kappa\tan\kappa=-1$ with multiple solutions 
$\kappa_1=2.798..$, $\kappa_2=6.121..$, etc.,  and asymptotically $\kappa_n=n\pi$ for integer $n$. 
In addition there is one negative energy solution for $n=0$ with $\kappa_0=1.997..$ 
which corresponds to the BCS-BEC crossover when $a\to -\infty$.
Generally, $n=0,1,2,..$ is the number
of nodes in the Jastrow wave function and each determines a new universal limit with
universal parameters depending on the number of nodes. The phase in the wave function
is $kb=\pi(n-1/2)$ whenever the unitarity limit of $n$ nodes is encountered.

\begin{figure}
\includegraphics[scale=0.6,angle=0]{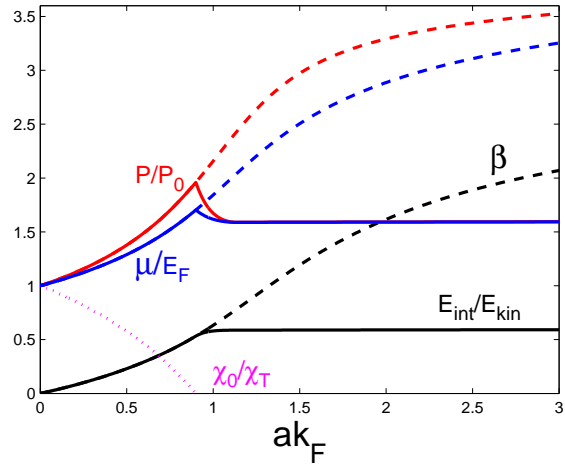}
\vspace{-0.5cm}
\caption{Universal functions calculated within JS at zero temperature vs. repulsive interaction: 
the ratio of interaction and kinetic energy $\beta$, the pressure and chemical potential
and the spin susceptibility $\chi_0/\chi_T$, all with respect to their non-interactive values. 
Full curves include the FM transition whereas the dashed have FM suppressed, i.e. remain in the PM phase.}
\end{figure}

It should be emphasised that for positive scattering lengths the wave
function and thus the correlations function between fermions of unlike
spin and bosons $\propto r\phi\sim\sin(kr-b)$ has a node at $b/k$ which is somewhere within
the interparticle distance. It does not vanish as $r\to0$
as does the wave function for a short range repulsive potential as in
hard sphere scattering where $a\simeq R$. Therefore the Gutzwiller
approximation may well apply for hard sphere gases, strongly
correlated nuclear fluids and liquid helium as discussed in
\cite{Zhai} but it does not apply to the repulsive unitarity limit of
ultracold gases when the wave function has to obey the short range
boundary condition of Eq. (\ref{a}).

It is customary to define the universal function
$\beta=E_{int}/E_{kin}$ as the ratio of the interaction $E_{int}$ and kinetic
energy $E_{kin}=(3/5)E_F$. In the JS model the interaction energy per particle is
$E_{int}=\hbar^2k^2/2m=\kappa^2/2md^2$ and thus
$\beta=(5/3)\kappa^2/k_F^2d^2$ in the PM phase. In the FM phase the
spin densities differ and the ratio of the
average interaction to kinetic energy can be considerably lower than $\beta$ as shown in Fig. (2).

Because Eq. (\ref{pos}) has a string of solutions for a given
scattering length or $k_Fa$, $\kappa$ and $\beta$ are multivalued
functions which we distinguish by the index $n=0,1,2,...$ referring to
the number of nodes in the many-body wave function between any two
atoms \cite{long}.  $\beta_0$ has been studied extensively in the BCS-BEC
crossover and $\beta_1$ in the repulsive crossover
\cite{Bourdel,Gupta,Jo}. In the repulsive unitarity limit $n=1$ the
universal parameter is
$\beta_1(k_Fa\to\infty)=5\kappa_1^2/3(3\pi)^{2/3}\simeq 2.93$.  It has
recently been measured for a $^6$Li gas in two spin states \cite{Jo}.
The chemical potential in the optical trap almost doubles going from
the non-interacting to the unitarity limit. Since it scales as
$\mu\propto\sqrt{1+\beta_1(\infty)}$ we obtain $\beta_1(\infty)\sim 3$
compatible with JS.  In the following we concentrate on repulsive
interactions and use $\beta=\beta_1$.

\begin{figure}
\includegraphics[scale=0.6,angle=0]{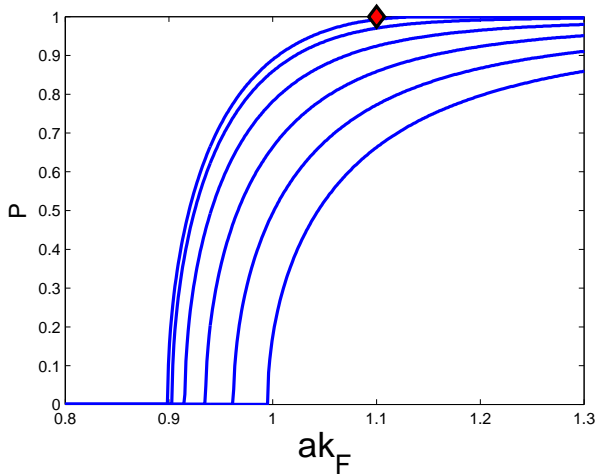}
\vspace{-0.5cm}
\caption{Polarization vs. $ak_F$ at $T/T_F=0,0.1,0.2,...,0.5$ from left to right. The second
order transition yields a steep but continuous transitions $P\propto\sqrt{ak_F-ak_c}$.
The diamond indicates the transition point to a pure one-component ($P=\pm 1$) FM at zero temperature.}
\end{figure}

The interaction energy for an atom with spin $\sigma$ depends on the density of
unlike spins and is given by
$\lambda=\kappa_{\sigma}^2/2md_\sigma^2\equiv (3/5)E_{F\sigma'}\beta(ak_{F,\sigma'})$,
where $\beta$ is the universal function for repulsive interactions.
We obtain the total energy density at zero temperature by adding the Fermi kinetic
energy and the interaction energy $\lambda_\sigma$, 
and sum over particle densities \cite{Vijay,long}
\begin{eqnarray} \label{LOCV+}
  f &=& \frac{3}{5}\sum_{\sigma} E_{F,\sigma}\rho_\sigma + 
    \frac{3}{5}\sum_{\sigma\ne\sigma'} E_{F,\sigma'}\beta(ak_{F,\sigma'})
           \rho_\sigma   \, +f_T \,,
\end{eqnarray}
including a thermal free energy $f_T$ as given above.  This expression
generalizes the standard expression for the energy density
$E/V=(3/5)E_F(1+\beta)\rho$ to finite polarization and temperature.
The result can be understood from dimensional arguments as $\beta$ is
dimensionless and gives the repulsive energy of particles of spin
$\sigma$ due interactions with particles of opposite spin.  Note that
the interaction energy and its dependence on polarization is given in
terms of one universal function $\beta$ of one variable only. As shown
in Fig. 2 the ratio of the interaction to kinetic energy is reduced by the FM transition w.r.t
$\beta$.

Expanding Eq. (\ref{LOCV+}) for small polarization gives
\bea \label{fJS}
  \frac{f}{f_0} &=& 1+\beta +\frac{5}{9}\frac{\chi_0}{\chi_T} P^2 +{\cal O}(P^4) \,,
\eea
where the isothermal spin susceptibility is
\bea \label{cJS}
  \frac{\chi_0}{\chi_{T}}=1-\frac{7}{5}\beta-\frac{2}{5} ak_F\beta'+\frac{1}{10}(ak_F)^2\beta''
  + \frac{\pi^2}{12}\frac{T^2}{T_F^2} \, ,
\eea
with $\beta'=d\beta/d(ak_F)$ and $\beta''=d^2\beta/d(ak_F)^2$.
In the dilute limit $\beta=(10/9\pi)ak_F$ and Eqs. (\ref{fJS}) and (\ref{cJS}) 
reduce to Eqs. (\ref{GL}) and (\ref{chiStoner}) respectively.

The spin-susceptibility calculated within JS is shown in Fig. 2 at zero temperature.
It diverges
at $ak_F\simeq0.90$ where the universal function is $\beta_{FM}\simeq0.53$.
By equating the energy of the unpolarized gas, $\sim(1+\beta)$ with that of
a fully polarized (one-component) gas, $\sim 2^{2/3}$, we find that a first order transition
requires $\beta=2^{2/3}-1\simeq0.59>\beta_{FM}$, and therefore
JS predicts a second order FM transition as shown in Fig. 1.
This transition point is in remarkable agreement with two recent QMC calculations which
find $ak_F=0.86$ \cite{Pilati} and $ak_F=0.89$ \cite{Chang}.
The QMC calculations could not determine the order of the transition within numerical accuracy. 
In the BCS-BEC crossover a minor discrepancy was found between JS \cite{long} 
and QMC \cite{Carlson,Casulleras}
which partly could be attributed to pairing which is excluded in the JS wave function. 
Since pairing is absent for repulsive interactions 
the JS model is expected to match the QMC calculations better near the FM transition.
Note that the JS wave function is also used as a starting point in the
QMC calculations of Refs. \cite{Chang,Carlson,Casulleras}.

Minimizing the free energy of Eq. (\ref{fJS}) we obtain the polarization
$P\propto\sqrt{-\chi_0/\chi_T}$ at the onset of FM as shown in Fig. 3 at low temperatures. Full polarization is reached at $ak_F\simeq 1.1$ at zero temperature only.

The spin-susceptibility is related to the spin-antisymmetric Landau parameter as
$F_0^A=(m^*/m)\chi_0/\chi_T-1$.
The effective mass $m^*=m$ is implicitly assumed in the JS energy of Eq. (\ref{LOCV+}).
It has recently been measured in the BCS-BEC unitarity limit $m^*_0/m=1.13\pm0.03$ 
\cite{Nascimbene} but not for the repulsive crossover yet. 
Since $\beta$ at the FM transition point is comparable to $|\beta_0|$,
these two effective masses may be expected to be similar. The small deviation from $m^*=m$ 
only changes the universal functions and the phase diagram slightly 
at higher temperature. The order and the position of the transition is unchanged at zero temperature.

\begin{figure}
\includegraphics[scale=0.6,angle=0]{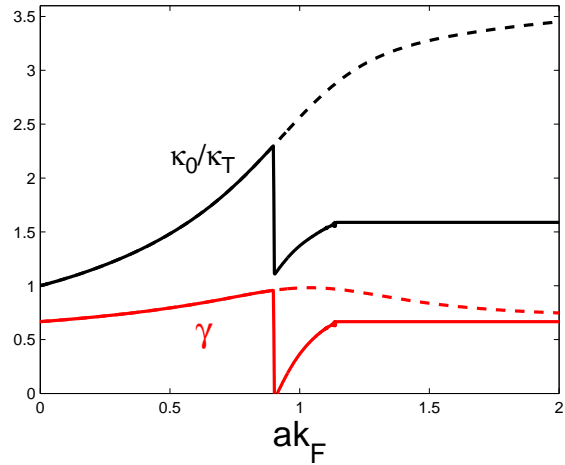}
\vspace{-0.5cm}
\caption{Compressibility and polytropic index vs. repulsive scattering length ($ak_F^0$) at zero 
temperature. Both are second derivates of the free energy and are therefore discontinuous at the
FM transition.} 
\end{figure}

\section{Number fluctuations}

The density or local number fluctuations have recently been measured
in shot noise experiments for an ideal ultracold Fermi gas and by
speckle noise in the BCS-BEC crossover \cite{Sanner,Muller}. The
number fluctuations are measured in a small subvolume of the atomic
cloud with almost uniform density. The fluctuations in spin and total
number of atoms are directly related to the spin susceptibility and
compressibility respectively.

The local fluctuations in total number can for a large number of atoms 
be related to the isothermal compressibility 
$\kappa_T=\rho^{-2} (\partial \rho/\partial \mu)_ {V,T}$, by
the fluctuation-dissipation theorem
\bea \label{Nfluc}
    \frac{(\Delta N)^2}{N}=\frac{3}{2}\frac{T}{T_F}\frac{\kappa_T}{\kappa_0} \, .
\eea
Here, $\kappa_0=3/(2\rho E_F)$ is the compressibility for an ideal Fermi gas at zero
temperatures. An ideal classical gas has $\kappa_T=1/(\rho k_BT)$ such that
the number fluctuations are Poisson: $(\Delta N)^2/N=1$.
The compressibility is related to the symmetric Landau parameter as
$F_0^S=(m^*/m)\kappa_0/\kappa_T-1$.

The compressibility can generally be expressed in terms of the universal function $\beta$ \cite{mode}
at zero temperature
\bea
 \frac{\kappa_0}{\kappa_T} = 1+\beta+\frac{4}{5}ak_F\beta'+\frac{1}{10}(ak_F)^2\beta'' ,
\eea 
in the PM phase. Once the FM transition sets in, the ground state energy of Eq. (\ref{LOCV+})
is lowered due to finite
polarization and the inverse compressibility drops as shown in Fig. 4 for JS.
It is discontinuous when the second order FM transition sets in 
because it is a second derivative of the free energy which is softened by the spin-susceptibility
term in Eq. (\ref{fJS}). In the pure one-component FM phase the
compressibility is that of an ideal one-component gas $\kappa_0/\kappa_T=2^{2/3}$.
The peculiar and discontinuous behaviour of the compressibility at the FM transition
is directly reflected in
the fluctuations in total number according to Eq. (\ref{Nfluc}).

If the FM transition was first order the compressibility diverges at
the phase transition, i.e., $\kappa_0/\kappa_T$ vanishes in part of
the density region where $0<P<1$ (see Figs. 3+4).  Consequently, the number
fluctuation also diverges according to
Eq. (\ref{Nfluc}) reflecting the density discontinuity at a first
order transition.

The fluctuation-dissipation theorem also relates the thermal spin fluctuations to the spin susceptibility
\bea 
 \frac{\Delta(N_\uparrow-N_\downarrow)^2}{N}
   = \frac{3}{2}\frac{T}{T_F}\frac{\chi_T}{\chi_0} \,. 
\eea
At the FM instability the spin-susceptibility and therefore also the spin fluctuations diverge
reflecting that phase separation occurs between domains of polarization $\pm P$. 
Such domains were, however, not observed in the experiments of \cite{Jo} 
within the spatial resolution of the experiment.

\section{Trap radii and kinetic energies}

In experiments the atoms are confined in harmonic traps. 
For a sufficiently large number $N=\sum_\sigma N_\sigma$ of particles confined in a (shallow) trap
the system size $R_\sigma$ is so long that
density variations and the extent of possible phase transition interfaces can be
ignored and one can apply the local density approximation.
The total chemical potential is given by the sum of the
harmonic trap potential and the local chemical potential 
$\mu_\sigma=(df/d\rho_\sigma)_{V,T}$
\bea \label{TF}
  \mu_\sigma(r) +\frac{1}{2}m\omega^2 r^2 
      = \frac{1}{2}m\omega^2 R_\sigma^2 \,,
\eea
which must be constant over the lattice for all components $\sigma=1,2,..,\nu$.
It can therefore be set to its value
at its edge $R_\sigma$, which gives the r.h.s. in Eq. (\ref{TF}). 
The equation of state determines the chemical potentials $\mu_\sigma$ in terms of
the universal function of Eq. (\ref{LOCV+}). 

In a two-component spin-balanced system the chemical potential and
radii of the two components are equal (denoted $\mu$ and $R$ in the following). In the FM
phase their densities $\rho(1\pm P)$ differ but these FM spin domains
coexist. Using the JS EoS of Eq. (\ref{LOCV+}) to calculate the
chemical potential we can find the density distribution from chemical
equilibrium Eq. (\ref{TF}) including phase transitions and calculate
cloud radii $R$, the root mean square $RMS=\sqrt{\av{r^2}}$ and
kinetic energy $E_{kin}=\av{k_F^2/2m}$ averaged over all particles in
the trap.  These are shown in Fig. 5 normalized to their values
trapped non-interacting ultracold atoms, $R_0=(24N)^{1/6}a_0$ ,
$RMS_0=\sqrt{3/8}R_0$ and $E_{kin}^0=(3/8)E_F^0$ respectively.  Here,
$E_F^0=(\hbar k_F^0)^2/2m$ and $k_F^0=(24N)^{1/6}/a_0$ are the Fermi
energy and wave number in the centre of the trap for non-interacting
atoms and $a_0=\sqrt{\hbar/m\omega}$ is the oscillator length.
Repulsive interactions reduce the central density and Fermi energy as
can be seen from $k_F/k_F^0$ shown in Fig. 5. As a consequence the
radii increase except for the RMS radius above
the FM transition (see Fig. 5). It decreases because atoms are redistributed from
the PM phase near the surface to the FM phase in the centre. The
kinetic energy of the atoms has the opposite behaviour because
repulsion increases the interaction energy in the PM phase at the cost
of the kinetic energy.

In the recent experiment of Ref. \cite{Jo} a transition is observed around $ak_F^0\simeq
2.2$ at temperatures $T/E_F^0=0.12$ (and $ak_F^0\simeq 4.2$ at
$T/E_F=0.22$).  This transition point is a factor of $\sim
2$ larger than the FM transition point calculated in
all models \cite{Stoner,Duine,Conduit,Pilati,Chang} as well as JS. Rescaling
$ak_F^0$ by a factor ~2 we find very good quantitative and qualitative
agreement with the data of \cite{Jo} as was found in the second order
calculation of Ref. \cite{Conduit}).  The distinct transitions in the
radii, kinetic energies and atomic losses are well reproduced.

\begin{figure}
\includegraphics[scale=0.6,angle=0]{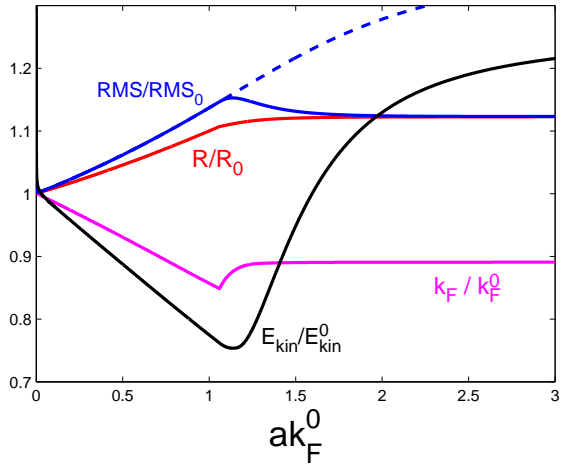}
\vspace{-0.5cm}
\caption{Radius of the trapped cloud, RMS radius, kinetic energy and central Fermi wavenumber 
all at zero temperature and relative to their non-interaction values
vs. repulsive scattering length ($ak_F^0$). Dashed curved shows the RMS radius for a PM phase
where the FM transition is inhibited. Due to repulsion the central density and thus $k_F$ is lower.} 
\end{figure}

However, the FM domains were not observed in Ref. \cite{Jo} within the spatial
resolution of the experiment.  Direct observation
should be possible in unbalanced spin systems where macroscopic FM
domain sizes can be realized.  As the repulsion is increased the RMS
radii of the minority spins increases faster than that of the majority
spins. When the FM transition occurs the system favours a core with
predominantly majority spins surrounded by a mantle of both spins in a
PM phase. With increasing spin imbalance the majority spin purity of
the FM core increases, i.e. the domains are effectively separated on a large scale. The
amount of separation and change in radii will depend on the overall
spin imbalance.

Three component systems with more than one Feshbach resonance as in
$^6$Li are also more complicated. For example, when the Feshbach
magnetic field is such that two resonances $a_{12}$ and $a_{13}$ are
large but $a_{23}$ small, the atoms will separate between a FM phase
of 1 and a mixed FM phase of 2+3 with different densities.

\section{Collective modes}

Collective modes have been studied intensively in the BCS-BEC crossover where they 
reveal important information of the equation of state (EoS) and determine $\beta_0(ak_F)$.
When the EoS can be approximated by a simple polytrope $P\propto \rho^{\gamma+1}$ the
collective eigen-frequencies can be calculated analytically \cite{Stringari,mode} in terms 
of the polytropic index $\gamma$. 
Even when the EoS is not a perfect polytrope the collective modes in the BCS-BEC crossover
could be described well using the effective polytropic index at densities near the
centre of the trap given by the logarithmic derivative \cite{mode}
\bea \label{gammaF}
    \gamma \equiv \frac{\rho}{P}\frac{dP}{d\rho} \, -1\, 
    =  \frac{\frac{2}{3}(1+\beta) +\frac{5}{6}ak_F\beta'
     +\frac{1}{6}(ak_F)^2\beta''}{1+\beta+ak_F\beta'/2} \,.
\eea
We therefore calculate $\gamma$ within JS for repulsive interactions
as shown in Fig. 4. Like the compressibility it has a discontinuity at the FM phase transition 
because it is a second derivative of the free energy with a second order transition.
In both the dilute limit and pure FM phase the gas is ideal with polytropic index $\gamma=2/3$.

For a very elongated or cigar-shaped trap (prolate in nuclear terminology), 
$\lambda\ll 1$, used in most experiments \cite{Thomas,Innsbruck}, 
the collective breathing modes separate into a low frequency axial mode with oscillation frequency
$\omega_{ax} = \sqrt{3-(\gamma+1)^{-1}}\, \omega_3$ and a
radial mode with $\omega_{rad} = \sqrt{2(\gamma+1)} \,\omega_0$ \cite{Stringari}.

The spin dipole mode is more complicated because it is 
sensitive to the spin susceptibility which diverges at the FM transition point. 
The EoS is far from polytropic and the delicate calculation of spin dipole modes with
diverging spin susceptibility is beyond the scope of this work. 
The spin dipole mode is estimated within a sum rule approach in Ref. \cite{Recati}.

\section{Multicomponent systems}

Interesting information on the order of the FM transition can be obtained by
generalizing the above results to Fermi gases with more that two
spin states such as $^6$Li with $\nu=3$ hyperfine states \cite{Jochim}, $^{137}$Yb with six nuclear
spin states \cite{Kitagawa}, and heteronuclear mixtures of $^{40}$K and $^6$Li \cite{Grimm}. 
The interactions and phases can be very complicated
when the Feshbach resonances between various components differ as for $^6$Li.
In the following we restrict ourselves to multi-components with the same relative scattering
length $a$.

In the dilute case the condition for a first order phase transition in a $\nu$ component system can be found from the energy density of Eq. (\ref{dilute}).
The preferred transition is directly from $\nu=1$ to a domains of one-components system $\nu=1$ which occurs when
\bea \label{crit}
  ak_F=\frac{9\pi}{10} \frac{\nu^{2/3}-1}{\nu-1} \left( 1 +
       \frac{5\pi^2}{12}\frac{T^2}{T_F^2}\nu^{-2/3}\right) .
\eea
At zero temperature this condition is $ak_F\simeq1.66, 1.53, 1.43,
1.36$, etc. for $\nu=2,3,4,5,6,..$ respectively.  The condition for a
second order transition is found by expanding the dilute
multi-component free energy for small polarization. One finds the same
spin susceptibility as in the two-component case,
Eq. (\ref{adilute}), and therefore the putative the second order
transition remains at $ak_F=\pi/2\simeq 1.57$.  Comparing numbers we
conclude that at zero temperature the second order transition occurs
for $\nu=2$ only in the dilute case whereas for $\nu\ge3$ the FM
transition is first order and given by Eq. (\ref{crit}).  At finite
temperatures the second order transition of Eq. (\ref{chiStoner})
match the first order of Eq. (\ref{crit}) at a temperature which
determines the tri-critical point $(ak_F,T)$ in the phase diagram for
$\nu\ge3$ as shown in Fig. 6.

The free energy of the JS model, Eq. (\ref{LOCV+}), also applies to
multi-component systems.  The condition for a first order FM
transition to coexisting fully polarized (one-component) FM domains is
\bea \label{betacrit}
  \beta(ak_F)= \frac{\nu^{2/3}-1}{\nu-1}\left( 1 +
       \frac{5\pi^2}{12}\frac{T^2}{T_{F}^2}
        \frac{m^*}{m}\nu^{-2/3} \right) .
\eea
At zero temperature the FM transition occurs for $\beta=0.59, 0.54,
0.51, 0.48,..$ at $ak_F=0.96, 0.91, 0.87, 0.84,..$ for
$\nu=2,3,4,5,..$ respectively.  As in the dilute case the
spin-susceptibility is unchanged, Eq. (\ref{cJS}), in the JS model and
the putative second order transition remains when $\beta=0.53$ at
$ak_F=0.90$.  Thus the FM transition at is at zero temperature
marginally second order for $\nu\le3$ but first order for $\nu\ge4$.
Again the tri-critical points $(ak_F,T)$ are determined by the matching
condition for the first Eq. (\ref{betacrit}) and second
Eq. (\ref{cJS}) order transitions and are shown in Fig. 6. 

Generally
the difference between first and second order FM transition is small
which may explain why QMC could not determine the order within numerical
accuracy \cite{Pilati,Chang}.
First order transitions to partially polarized FM does not occur for two-component
systems but may be possible in multi-component systems.

The marginal first vs. second order FM transition for $\nu=3$ is
analogous to the marginal stability in the unitary limit of the
BCS-BEC crossover \cite{long}. Here it is known that two-component
systems are stable but four-component systems are unstable as in
nuclear matter.

\begin{figure}
\includegraphics[scale=0.6,angle=0]{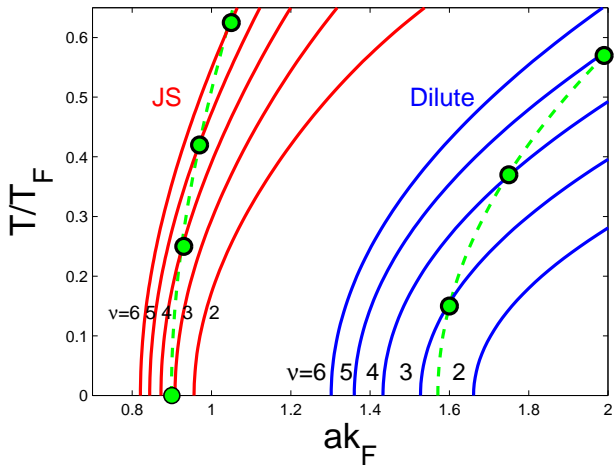}
\vspace{-0.5cm}
\caption{Phase diagrams for multi-component ($\nu=2,3,4,5,6$ from left to right) 
Fermi gases with repulsive interactions.
Full (dashed) curves indicate first (second) order PM to FM 
transitions within JS and dilute approximations. 
Circles indicate the tri-critical points where the transition changes from first to second order at 
higher temperatures.}
\end{figure}

\section{Summary and outlook} 

By extending the Jastrow-Slater approximation to finite polarization
and temperature we have calculated a number of thermodynamic functions
and observables for cold Fermi atoms with repulsive interactions. In
particular we found a second order FM phase transition at
$ak_F\simeq0.90$ at zero temperature in close agreement with QMC. 
The compressibility and spin susceptibility were calculated and the resulting observables like the
fluctuations in total number and spin as well as collective modes are
discontinuous at the transition point. 
These can be distinguished from a first order transition where, e.g., the 
compressibility diverges.
 
For trapped gases the radii and kinetic energies also have characteristic behaviour
as function of repulsive interaction strength when the FM transition occurs in the centre. 
If the interaction strength is reduced by a factor $\sim 2$ the radii
and kinetic energies of JS and Ref. \cite{Conduit} agree qualitatively and quantitatively
with experiments \cite{Jo}. In order to observe the FM domains we suggest to start out with a
spin-imbalanced system of two-component Fermi atoms and tune the
magnetic field towards the Feshbach resonance from the repulse side
where the FM transition sets in. As result the core will be a large
domain of the majority spin only which exceeds the experimental domain
size resolution.  

It would be interesting to study multi-component systems such as the three
component $^6$Li system near Feshbach resonances where bulk
separation between the spin component domains is predicted to take place.
Multi-component systems with the same interactions (and scattering lengths) between
states display interesting phase diagrams with first to second order tri-critical points
when the number of components exceeds two in the dilute case and three in the JS model.


\end{document}